\documentstyle{mn}

\newif\ifAMStwofonts
\ifoldfss
  \ifCUPmtlplainloaded \else
    \NewTextAlphabet{textbfit} {cmbxti10} {}
    \NewTextAlphabet{textbfss} {cmssbx10} {}
    \NewMathAlphabet{mathbfit} {cmbxti10} {} % for math mode
    \NewMathAlphabet{mathbfss} {cmssbx10} {} %  "   "    "
  \fi
  \ifAMStwofonts
    \ifCUPmtlplainloaded \else
      \NewSymbolFont{upmath} {eurm10}
      \NewSymbolFont{AMSa} {msam10}
      \NewMathSymbol{\upi}     {0}{upmath}{19}
      \NewMathSymbol{\umu}     {0}{upmath}{16}
      \NewMathSymbol{\upartial}{0}{upmath}{40}
      \NewMathSymbol{\leqslant}{3}{AMSa}{36}
      \NewMathSymbol{\geqslant}{3}{AMSa}{3E}

      \let\leq=\leqslant 
      \let\geq=\geqslant 
    \fi
  \fi
\fi % End of OFSS

\ifnfssone
  \newmathalphabet{\mathit}
  \addtoversion{normal}{\mathit}{cmr}{m}{it}
  \addtoversion{bold}{\mathit}{cmr}{bx}{it}
  \newmathalphabet{\mathbfit} % math mode version of \textbfit{..}
  \addtoversion{normal}{\mathbfit}{cmr}{bx}{it}
  \addtoversion{bold}{\mathbfit}{cmr}{bx}{it}
  \newmathalphabet{\mathbfss} % math mode version of \textbfss{..}
  \addtoversion{normal}{\mathbfss}{cmss}{bx}{n}
  \addtoversion{bold}{\mathbfss}{cmss}{bx}{n}
  \ifAMStwofonts
    \ifCUPmtlplainloaded \else
      \UseAMStwoboldmath
      \makeatletter
      \new@mathgroup\upmath@group
      \define@mathgroup\mv@normal\upmath@group{eur}{m}{n}
      \define@mathgroup\mv@bold\upmath@group{eur}{b}{n}
      \edef\UPM{\hexnumber\upmath@group}
      \new@mathgroup\amsa@group
      \define@mathgroup\mv@normal\amsa@group{msa}{m}{n}
      \define@mathgroup\mv@bold\amsa@group{msa}{m}{n}
      \edef\AMSa{\hexnumber\amsa@group}
      \makeatother
      \mathchardef\upi="0\UPM19
      \mathchardef\umu="0\UPM16
      \mathchardef\upartial="0\UPM40
      \mathchardef\leqslant="3\AMSa36
      \mathchardef\geqslant="3\AMSa3E

      \let\leq=\leqslant 
      \let\geq=\geqslant 
    \fi
  \fi
\fi % End of NFSS release 1

\ifnfsstwo
  \DeclareMathAlphabet{\mathbfit}{OT1}{cmr}{bx}{it}
  \SetMathAlphabet\mathbfit{bold}{OT1}{cmr}{bx}{it}
  \DeclareMathAlphabet{\mathbfss}{OT1}{cmss}{bx}{n}
  \SetMathAlphabet\mathbfss{bold}{OT1}{cmss}{bx}{n}
  \ifAMStwofonts
    \ifCUPmtlplainloaded \else
      \DeclareSymbolFont{UPM}{U}{eur}{m}{n}
      \SetSymbolFont{UPM}{bold}{U}{eur}{b}{n}
      \DeclareSymbolFont{AMSa}{U}{msa}{m}{n}
      \DeclareMathSymbol{\upi}{0}{UPM}{"19}
      \DeclareMathSymbol{\umu}{0}{UPM}{"16}
      \DeclareMathSymbol{\upartial}{0}{UPM}{"40}
      \DeclareMathSymbol{\leqslant}{3}{AMSa}{"36}
      \DeclareMathSymbol{\geqslant}{3}{AMSa}{"3E}

      \let\leq=\leqslant 
      \let\geq=\geqslant 
    \fi
  \fi
\fi % End of NFSS release 2

\ifCUPmtlplainloaded \else
  \ifAMStwofonts \else % If no AMS fonts
    \def\upi{\pi}
    \def\umu{\mu}
    \def\upartial{\partial}
  \fi
\fi

\newcommand{\beq}{\begin{equation}}
\newcommand{\beqa}{\begin{eqnarray}}
\newcommand{\eeq}{\end{equation}}
\newcommand{\eeqa}{\end{eqnarray}}

\title{Cosmology with x-matter
}
\author[T.Chiba et al.]{Takeshi Chiba,$^1$ Naoshi Sugiyama$^2$ and 
Takashi Nakamura$^1$\\
$^1$Yukawa Institute for Theoretical Physics, Kyoto University, 
Kyoto 606-01, Japan\\
$^2$Department of Physics, Kyoto University, 
Kyoto 606-01, Japan}

\date{Accepted .
      Received ;
      in original form}

\pagerange{\pageref{firstpage}--\pageref{lastpage}}
\pubyear{1997}

\begin{document}

\maketitle

\label{firstpage}

\begin{abstract}
Motivated by the possibility of  $H_0t_0 > 1$ where $H_0$ and $t_0$ are the Hubble parameter and
the age of the universe, respectively, 
we investigate the cosmology including x-matter.
x-matter is expressed by  the equation
of state $p_x=w_0\rho_{x0} + c_s^2(\rho_x-\rho_{x0})$,
where $p_x$, $\rho_x$ and $\rho_{x0}$ are the pressure, the density 
of x-matter and the density at present, respectively. 
$w_0$ and $c_s^2$ are functions of $\rho_x$ in general.
 x-matter has the most general 
form of the equation of state which is characterized by 
1)violation of strong energy condition  at present
for $w_0 < -1/3$ ; 2) locally stable (i.e. $c_s^2\geq 0$); 3)causality
is guaranteed ($c_s\leq 1$).  Considering the case that
$w_0$ and $c_s^2$ are constants, we  find that there is a large
parameter space of $(w_0,c_s^2,\Omega_{x0})$ in which the model
universe is consistent with the age of the universe and the
observations of distant Type I  supernovae.  

\end{abstract}


\section{introduction}
The age of the globular clusters $15\pm 2{\rm
Gyr}$ \cite{gc} and the age of the universe with current $H_0 = 70\pm
10$km/s/Mpc (Freedman 1996; Riess et al. 1995,1996) may suggest
  $H_{0}t_0 > 1$ where $H_0$ and $t_0$
are the Hubble parameter and the age of the universe, respectively.\footnote{
The recent Hipparcos results(Feast \& Catchpole 1997; Reid
1997)suggest the globular cluster age 
$12 {\rm Gyr}$ and $H_0 \simeq 65$km/s/Mpc, and then $H_{0}t_0 \ga
0.82$. The age problem, however, still persists 
if $\Omega_{M0}\ga 0.27$.}
$H_{0} t_0 > 1$ stands for the apparent contradiction of the age of the 
universe; the universe is younger than the oldest globular cluster.
For this age problem   
 Nakamura et al. \shortcite{nncs} proved the theorem such that
 if $H_{0} t_0 > 1$, either
1) Einstein's theory of gravity is not correct theory of gravitation, 
2) the strong energy condition is not satisfied, or 
3) foliation by geodesic slicing does not exist.

For the age problem of the universe, usually the existence of 
the cosmological term is suggested. However, recent SNIa survey by Perlmutter et al. \shortcite{sn} and
the statistics of gravitational lensing \cite{koch} are giving some
doubts on the $\Lambda$-dominated models.  From the point view of the
above general theorem \cite{nncs} the constant cosmological term is 
nothing but an example to save the age problem.  
For instance, the model with the  scalar field  which works as a 
time varying cosmological term was proposed based on particle physics 
(Fujii 1989) and investigated 
the role in cosmology(Peebles \& Ratra 1988; 
Sato, Terasawa \& Yokoyama 1989; Sugiyama \& Sato 1992).
Recently, Steinhardt \shortcite{st} 
suggested an alternative to lambda term in which a
matter with $p=w\rho(-1 < w < 0)$ resides in addition to
ordinary CDM.

Turner and White\shortcite{wt} have done comprehensive study 
of this so-called xCDM model. 
In their paper, they mentioned that 
they do not allow for clumping of xCDM because a component with 
$w <0$ is highly unstable to growth of perturbations on small scales.
Therefore they assumed xCDM to be a smooth component on small scales.
They suggested kinds of 
topological defects may behave like  such ``xCDM''.  
However, $w \equiv p/\rho < 0 $ does not necessarily imply 
$c_s^2 \equiv \delta p / \delta\rho < 0$ which causes the instability. 
Thus we may construct a clumpy xCDM model such that 
$w < 0$ {\it and} $c_s^2 \geq 0$ if we can find 
proper energy momentum tensor and equation of state.
As shown on pp.89 of Hawking and Ellis \shortcite{he}, 
the energy momentum tensor is
classified into four types. The form of the energy momentum
tensor compatible with the global isotropy of the universe should be
of Type I with pressure $p_1=p_2=p_3=p$.
In this letter, we shall study the general cosmology, in which
x-matter  with $w < 0$ and $c_s^2 \geq 0$
resides in addition to ordinary cold dark matter (CDM), 
baryonic matter and radiation.
This kind of general study may 
give us some insights into the age problem.

\section{basics}

As mentioned in \S 1, x-matter should possess following nature, i.e., 
$w \equiv p_x/\rho_x < -1/3$ to violate strong energy condition and 
$c_s^2 \equiv \delta p /\delta \rho \geq 0$ to stabilize growth of 
perturbations. The former condition is necessary (though not sufficient) 
to elongate the cosmic
age \cite{nncs}. Several matter fields are known to violate the strong
energy condition: a massive scalar field, a domain wall and the
cosmological constant.  
$c_s$ represents, as shown, the sound velocity of x-matter and
determines how the density evolves. 
Together with the latter condition,
we demand that it is not faster than the light velocity.  
As the simplest form, we employ the equation of state of
x-matter as 
\beq p_x=w_0\rho_{x0}+c_s^2(\rho_x-\rho_{x0}), 
\eeq 
where
the suffix $0$ denotes the present value.
Although $w_0$ and
$c_s$ are functions of $\rho_x$ in general, to simplify the discussion
we assume that they are constants in this letter.  
Our x-matter model includes the cosmological constant model 
as a special case since $p_x=-\rho_{x0}$ and
$\rho_x=\rho_{x0}$ for $w_0=-1$. 
On the other hand, xCDM by Turner and White\shortcite{wt} is 
not included since in xCDM
$c_s^2 < 0$ so that it is unstable for all scale perturbations. 
According to above requirements, we consider the region $-1\leq w_0 < -1/3$, 
$0 \leq c_s^2 \leq 1$, and $\rho_x \geq 0$. 
A sketch of the equation of state 
is shown in Fig.1.

The energy equation is 
\beq
{d \rho_x\over d a}=-{3\over a}(\rho_x+p_x) ,
\eeq
where $a$ is the scale factor. Normalizing $a$ as  $a_0=1$, we have 
\beq
\rho_x={\rho_{x0}\over 1+c_s^2}\left
( (1+w_0)a^{-3(1+c_s^2)}+c_s^2-w_0\right).
\label{rho}
\eeq
We note that the cosmological constant model is an asymptotic limit in
our x-matter model: $p_x/\rho_x \rightarrow -1$ as
$a\rightarrow \infty$. 

The Friedmann equation is
\beqa
{1\over H_0^2}\left({\dot{a}\over
a}\right)^2 &+& {\Omega_{M0}+\Omega_{x0}-1\over a^2}\nonumber\\
&=&{\Omega_{M0}\over a^3} + {\Omega_{r0}\over a^4}  \label{frw} \\
&&+{\Omega_{x0}\over 1+c_s^2}\left 
( (1+w_0)a^{-3(1+c_s^2)}+c_s^2-w_0\right), \nonumber 
\eeqa
where $\Omega_{M0}$, $\Omega_{r0}$ and $\Omega_{x0}$ are the present 
density parameter of CDM, radiation (photon and neutrino)
and x-matter, respectively and over dot denotes 
time derivative.

Let us consider first the  constraint on $c_s^2$. This comes
from the requirement that  
the process of Big Bang Nucleosynthesis (BBN) is not appreciably 
disturbed  by the existence of x-matter.
Eq.(\ref{rho}) shows that 
x-matter grows  faster than dust matter as $a$ becomes smaller.
If $c_s^2 > 1/3$, it does even faster
than radiation and would dominate over radiation.
In order that the density of x-matter is smaller than 
the one of a single massless neutrino species at BBN 
($T_{\rm BBN} \simeq 0.1\rm MeV$), 
\beq
\Omega_{x0}{1+w_0\over 1+c_s^2} < 1.1 \times 10^{-5} 
\left(0.7\over h \right)^2 a_{{\rm BBN}}^{-1+3c_s^2} ,
\eeq
where $a_{{\rm BBN}} = 2.3\times 10^{-9} (T_{\rm BBN} / 0.1 {\rm MeV})$
is the scale factor at BBN and $h$ is the non-dimensional 
Hubble parameter normalized by $100\rm km/s/Mpc$.
Constraints on the $w_0-c_s^2$ plane are shown in Fig.2.
   From this figure, we can conclude that
$c_s^2 \la 0.15$ should be required.
 
The typical evolution of densities are shown in
Fig.3 for $c_s^2=0$ and $0.2$.  In the early universe x-matter is 
essentially the same as CDM for $c_s^2=0$ but exceeds CDM for $c_s^2 > 0$. 
On the other hand, for $a > 0.1$ where  the dominant contribution to
the age of the universe is made, it is different from CDM regardless of 
the value of $c_s^2$ as is shown in Fig.3.

\section{cosmology}

Now let us develop the cosmology with x-matter. 
Hereafter we consider the universe during the matter-dominated era.

\subsection{Critical Values}
Likewise the cosmology with the cosmological constant, there are some
critical values of $\Omega_{x0}$ which correspond to the boundary
between the  universe with or without the big-bang; 
the boundary between the recollapsing universe and the 
expanding universe. 

The Friedmann equation (\ref{frw}) 
can be read as the ``energy conservation equation'', 
the kinetic energy term being $\dot{a}^2/H_0^2$, the potential energy 
term $V(a)$ being 
\beq
V(a)=-{\Omega_{M0}}a^{-1}
-{\Omega_{x0}\over 1+c_s^2}\left
( (1+w_0)a^{-1-3c_s^2}+(c_s^2-w_0)a^2\right),
\eeq
and the total energy $E$ being $E=1-\Omega_{M0}-\Omega_{x0}$. 
The universe can recollapse or bounce if $E \leq max(V(a))$. For $c_s^2=0$,  
the maximum of $V(a)$ is evaluated analytically 
\beq
V(a)\leq -{3\over 2}
\left(\Omega_{M0}+(1+w_0)\Omega_{x0}\right)^{2/3}
(-2w_0\Omega_{x0})^{1/3}. 
\eeq
Therefore the condition $E \leq max(V(a))$ reads
\beq
4(1-\Omega_{M0}-\Omega_{x0})^3 \leq
27w_0\Omega_{x0}\left(\Omega_{M0}+(1+w_0)\Omega_{x0}\right)^2.
\label{cond}
\eeq
The equality, which is a cubic equation of $\Omega_{x0}$, 
 decides the critical values of $\Omega_{x0}$. 

Fig.4 shows the critical $\Omega_{xc1},\Omega_{xc2}$ for $c_s^2=0$ and 
several $w_0$. 
If $\Omega_{x0} > \Omega_{xc1}$, then the universe bounces without
the big-bang; if $\Omega_{xc2} < \Omega_{x0} < \Omega_{xc1}$, the
universe with big-bang expands forever; if $\Omega_{x0} < \Omega_{xc2}$, 
recollapses. The difference between bounce and recollapse comes from
the value of $a$ (=$a_m$)  at the maximum value of $V(a)$
(i.e. whether $a_m < 1$ or $a_m > 1$).

\subsection{Age of the Universe}

The cosmic age is given by 
\beqa
H_0t_0&=&\int_0^1
{da\over
  a}\Bigl[\Omega_{M0}a^{-3}+(1-\Omega_{M0}-\Omega_{x0})a^{-2} 
  \label{age}\\
&+&{\Omega_{x0}\over 1+c_s^2}\left
( (1+w_0)a^{-3(1+c_s^2)}+c_s^2-w_0\right)
\Bigr]^{-1/2}.\nonumber
\eeqa
The functional form of $H_0t_0$ shows that it is a decreasing function 
of $c_s^2$ or $w_0$. 
The current constraints on $h=0.70\pm
0.10$ (Freedman 1996; Riess et al. 1995,1996) 
and the globular cluster age 
$15\pm 2{\rm Gyr}$ \cite{gc} suggest at least $H_0t_0 \geq 0.80$. 
In Figs.5, the contour plots of $H_0t_0$ in
the $(\Omega_{M0},\Omega_{x0})$ plane are shown. 
For fixed $\Omega_{M0}$, $H_0t_0$ is an increasing function of 
$\Omega_{x0}$ in Figs.5(b), 5(d), 5(f) and 5(g), while it is a
decreasing one in Figs. 5(a) and 5(c). In Fig.5(e) 
it is either an increasing or a decreasing function, depending on
$\Omega_{M0}$. The reason for this behavior can be understood well 
by differentiating equation (\ref{age}) by $\Omega_{x0}$ as
\beqa
& &\frac {\partial H_0t_0}{\partial \Omega_{x0}} \nonumber \\
&=& \int_0^1
{da\over
  2a}\left(a^{-2}-{1\over 1+c_s^2}
( (1+w_0)a^{-3(1+c_s^2)}+c_s^2-w_0)\right) \nonumber \\
 &\times& \Bigl[ \Omega_{M0}a^{-3}+(1-\Omega_{M0}-\Omega_{x0})a^{-2}\nonumber \\
&+&{\Omega_{x0}\over 1+c_s^2}
\left( (1+w_0)a^{-3(1+c_s^2)}+c_s^2-w_0)\right)\Bigr]^{-3/2}.
\eeqa
For $w_0=-1$, the sign of $ {\partial H_0t_0}/{\partial \Omega_{x0}}$
is determined by the sign of $a^{-2}-1 >0$ so that 
$H_0t_0$ is an increasing function of $\Omega_{x0}$. This also  explains
the behaviors for $w_0=-0.8$ as follows. 
With the increase of $c_s$, the relative 
importance of the negative term in the expression of 
$ {\partial H_0t_0}/{\partial \Omega_{x0}}$ increases so that 
$ H_0t_0$ becomes a slowly increasing function of $\Omega_{x0}$.
As $c_s$ is increased, 
x-matter behaves  
like radiation so that $ H_0t_0$ should approach the value
 of the  radiation case. 
Note here $ H_0t_0=0.5$ in the flat radiation dominant universe.
For $w_0=-0.6$, on the other hand, 
$ H_0t_0$ is almost independent of $\Omega_{x0}$.
This shows that x-matter is ineffective to increase $ H_0t_0$
considerably  for $w_0 > -0.6$. 
It is easy to show that for $w_0 > -1/3$ x-matter does not
save the age problem at all, which is already proved by the general
theorem  \cite{nncs}.

\subsection{Luminosity Distance}
We  consider the luminosity distance-redshift relation in x-matter  models. 
The luminosity distance is defined by 
\beqa
H_0d_L(z)&=&
(1+z)\int_0^zdz'
\Bigl[\Omega_{M0}(1+z')^3\nonumber\\
&+&(1-\Omega_{M0}-\Omega_{x0})(1+z')^2\\
&+&{\Omega_{x0}\over
1+c_s^2}\left ( (1+w_0)(1+z')^{3(1+c_s^2)}+c_s^2-w_0\right)
\Bigr]^{-1/2}.\nonumber
\eeqa
Incidentally, the deceleration parameter $q_0$ is given by 
$q_0=\Omega_{M0}/2+(1+3w_0)\Omega_{x0}/2$.
The luminosity distance-redshift relation differs for two cosmological 
models with the same $q_0$ and $H_0$; it depends on the equation of
state of x-matter. 
The current uncertainty at $95\%(68\%)$ C.L. 
in the distance to  SNIa at $z\simeq 0.4$ 
is $0.38(0.40) < H_0d_L(z=0.4) < 0.51(0.48)$ 
(Perlmutter et al. 1995,1997; Turner \& White 1997).

In Fig.6  regions of $w_0$ and $\Omega_{x0}$ which satisfy both 
$H_0t_0 > 0.80$ and $0.38 < H_0d_L(z=0.4) < 0.51$ are shown.
 A region
bounded by a solid line is for  $\Omega_{M0}=0.1$, a dotted line for 
$\Omega_{M0}=0.3$, a dashed line for $\Omega_{M0}=0.5$.   
Upper lines come from the age limit which can be easily understood from 
the argument in the previous subsection. Lower right lines mean that
too much $\Omega_{x0}$ increases the luminosity distance above the 
observational limit $H_0d_L(z=0.4) < 0.51$.
We find that large parameter spaces are left even for a closed universe
model as well as an open universe model. 
The $\Lambda$-dominated universe 
corresponds to the $w_0=-1$ line in Fig.6. It is interesting to note
that high $\Omega_{M0}(> 0.3)$ cosmological models are allowed in
the x-matter model although excluded in the 
$\Lambda$-dominated universe.

\section{summary}
We have investigated the possibility that the dark matter component has 
the equation of state $p_x=w_0\rho_{x0}+c_s^2(\rho_x-\rho_{x0})$ such
that $-1\leq w_0 < -1/3$ and $0 \leq c_s^2 \leq 1$. 
We have studied various limits on $w_0$, $c_s^2$ and $\Omega_{x0}$. 
We have found that there is a large
parameter space of $(w_0,c_s^2,\Omega_{x0})$ in which the model
universe is consistent with the age of the universe and the
observations of distant Type I  supernovae.  
It may not be so meaningful to assess the ``best fit'' values of
$w_0,c_s^2,\Omega_{x0}$ because observational data will be updated
soon and the situation is not settled yet. 
It is interesting to study the properties of the large scale structure 
in x-matter models. 
We intend to extend a detailed study of perturbation theory and
``neoclassical '' tests of x-matter models. 
Those include the statistics of gravitational lensing and spectra of the 
cosmic microwave background radiation. 
The results will be reported in a future publication 
(Chiba, Sugiyama, and Nakamura, work in progress). 

\section*{Acknowledgments}
This work was supported in part by a
Grant-in-Aid for Basic Research of the Ministry of Education,
Culture, and Sports Nos.\ 08NP0801(TN) and 09440106(NS).

%%%%%%%%%%%%%%%%%%%%%%%%%%%%%%%%%%%%%%%%%%%%%%%%%%%%%%%%%%%%%%%%%%%%%%
\bsp
\section*{Figure Captions}

\vspace*{12pt}
\noindent
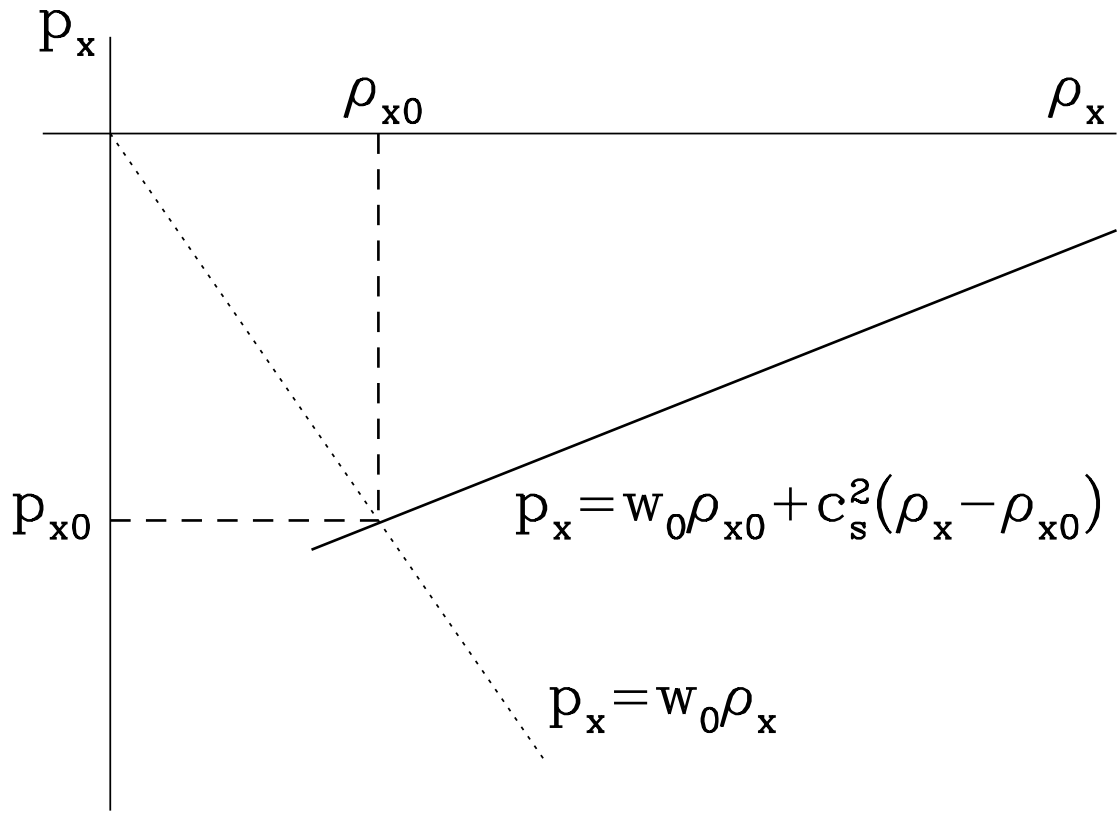
{\bf Figure 1: } A sketch of the equation of state of ``x-matter''. 
The solid line shows $p_x=w_0\rho_{x0}+c_s^2(\rho_x-\rho_{x0})$ and 
the dotted line shows $p_x=w_0\rho_{x0}$. 

\vspace*{12pt}
\noindent
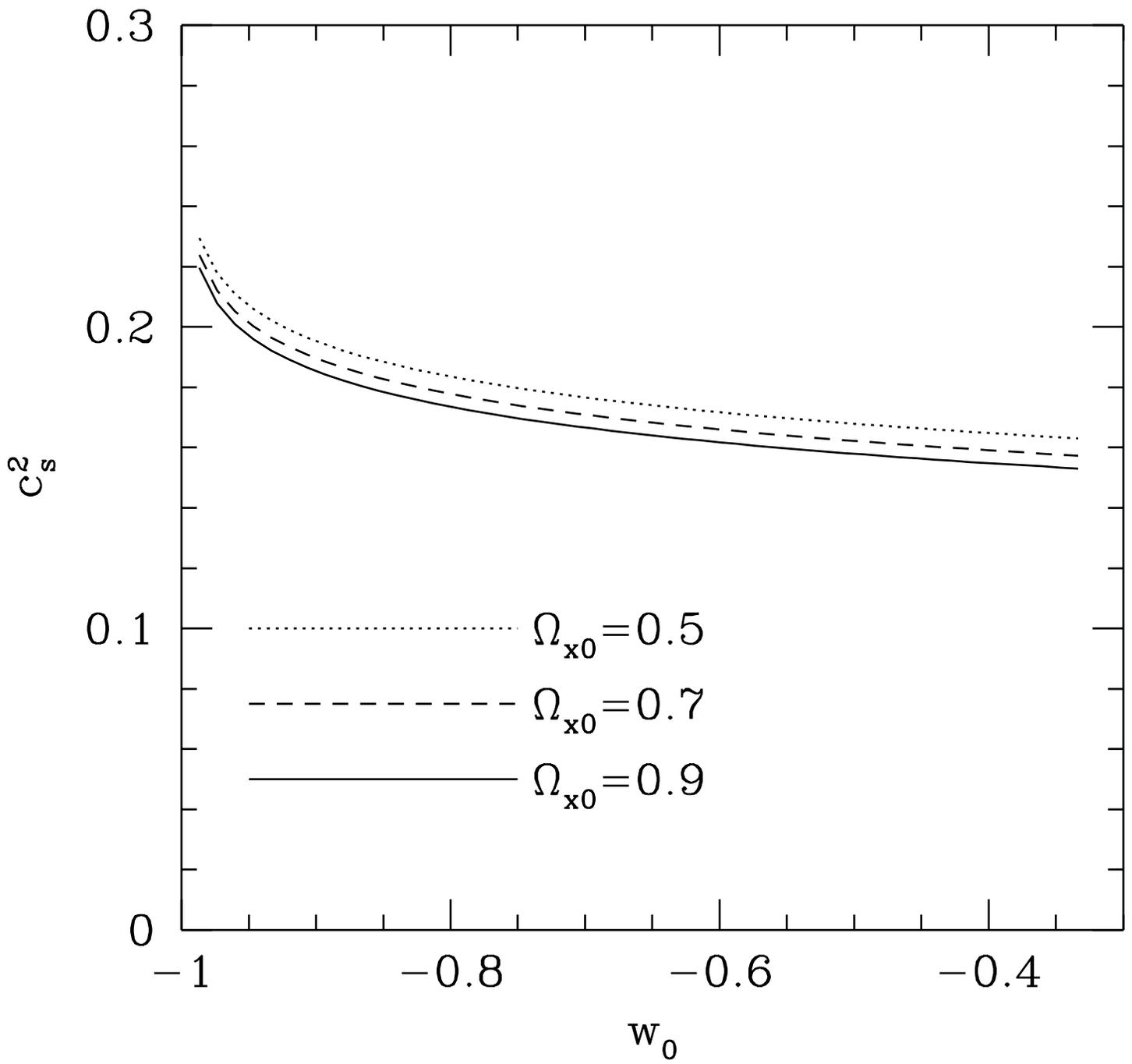
{\bf Figure 2: }Constraints from BBN on the $w_0-c_s^2$ plane.
The lines show that the energy density of x-matter is equal to the 
one of a single species neutrino for $\Omega_{x0}=0.5, 0.7$ and $0.9$.
Regions above these lines are excluded.  $T_{BBN}=0.1\rm MeV$ and
$h=0.7$ are employed.

\vspace*{12pt}
\noindent
{\bf Figure 3: } The evolution of densities $\rho_x,\rho_M$(matter density),
$\rho_{\gamma}$(radiation density) 
for $c_s=0$ and $0.2, w_0=-0.8, \Omega_{M0}=0.1$, 
and $\Omega_{x0}=0.9$. The scale factor is
normalized as $a=1$ at the present. 

\vspace*{12pt}
\noindent
{\bf Figure 4: } The critical values of $\Omega_{x0}$ for
$w_0=-1,-0.8,-0.6$ are shown.  Upper lines correspond to
$\Omega_{xc1}$ and  lower lines correspond to $\Omega_{xc2}$(see text 
for their definition). For $\Omega_{x0} > \Omega_{xc1}$,  the
universe bounces without the big-bang; for $\Omega_{xc2} < \Omega_{x0} <
\Omega_{xc1}$, the universe with big-bang expands forever;
$\Omega_{x0} < \Omega_{xc2}$, recollapses. 
 The difference between bounce and recollapse comes from
the value of $a$ (=$a_m$)  at the maximum value of $V(a)$
(i.e. whether $a_m < 1$ or $a_m > 1$).

\vspace*{12pt}
\noindent
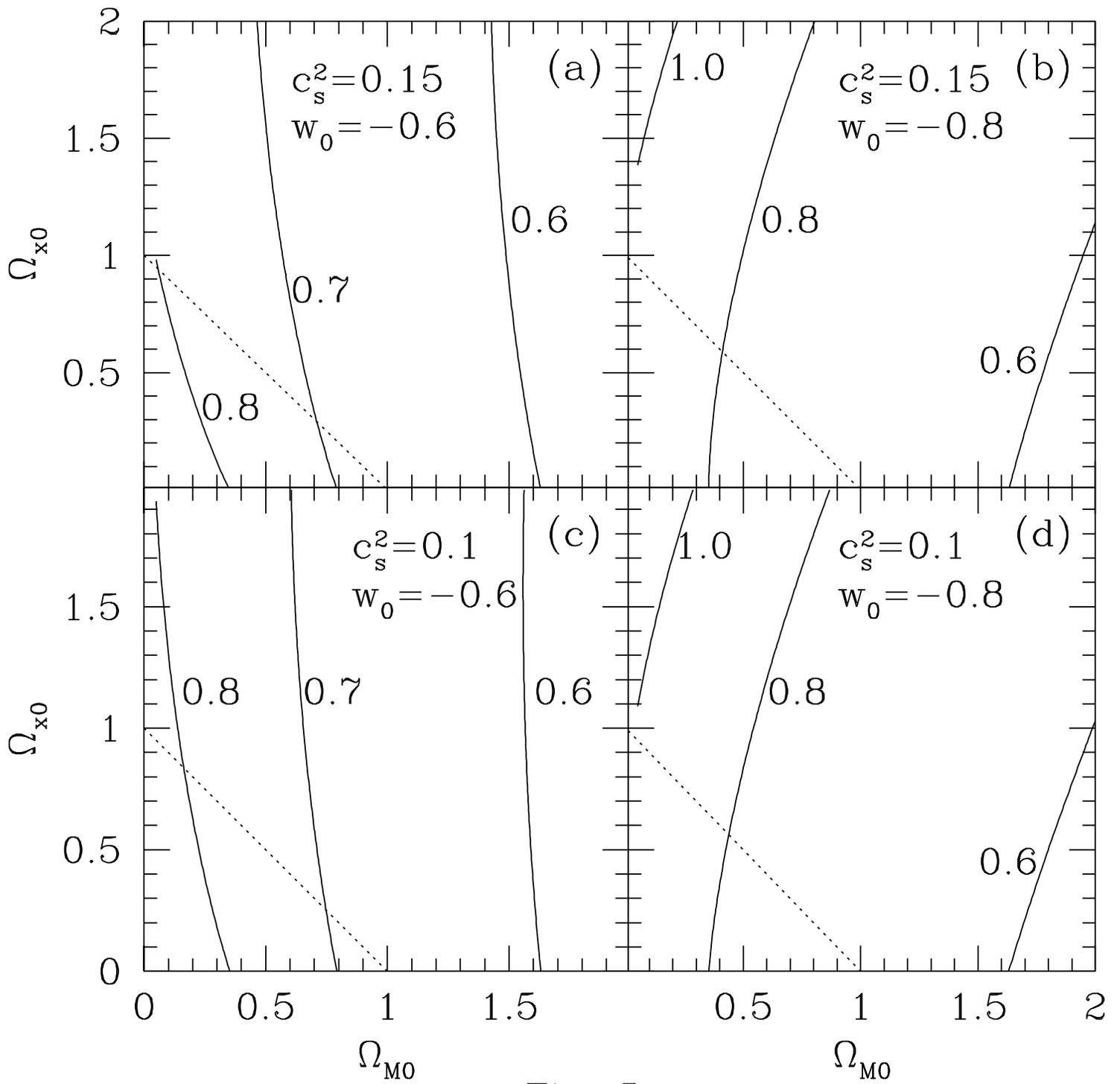
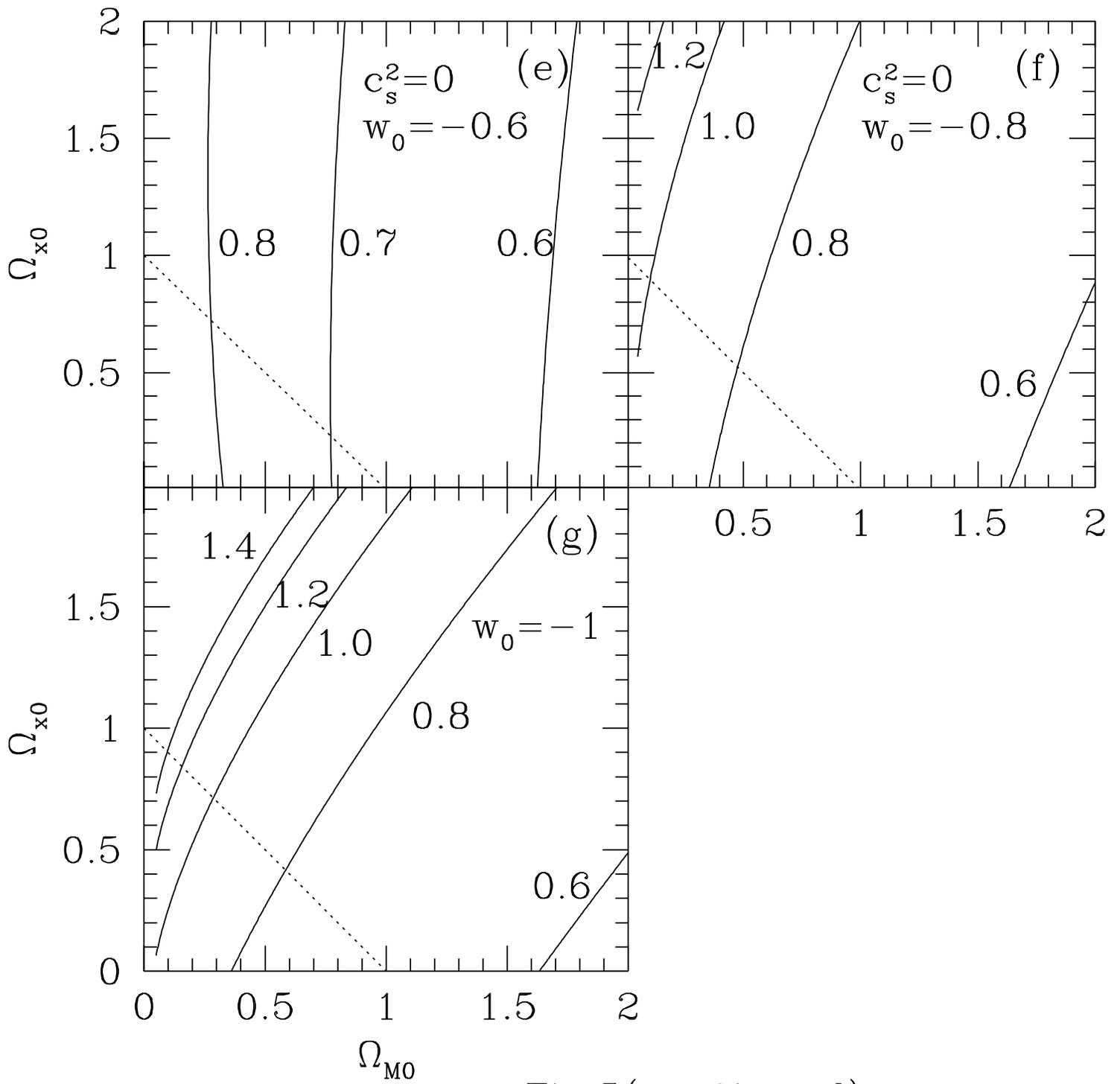
{\bf Figure 5: } The contour of $H_0t_0$ for $c_s^2=0,0.1,0.15$
and $w_0=-1,-0.8,-0.6$. 
A dotted line corresponds to a flat universe. 
Note that the $w_0=-1$ case is independent of $c_s$. 

\vspace*{12pt}
\noindent
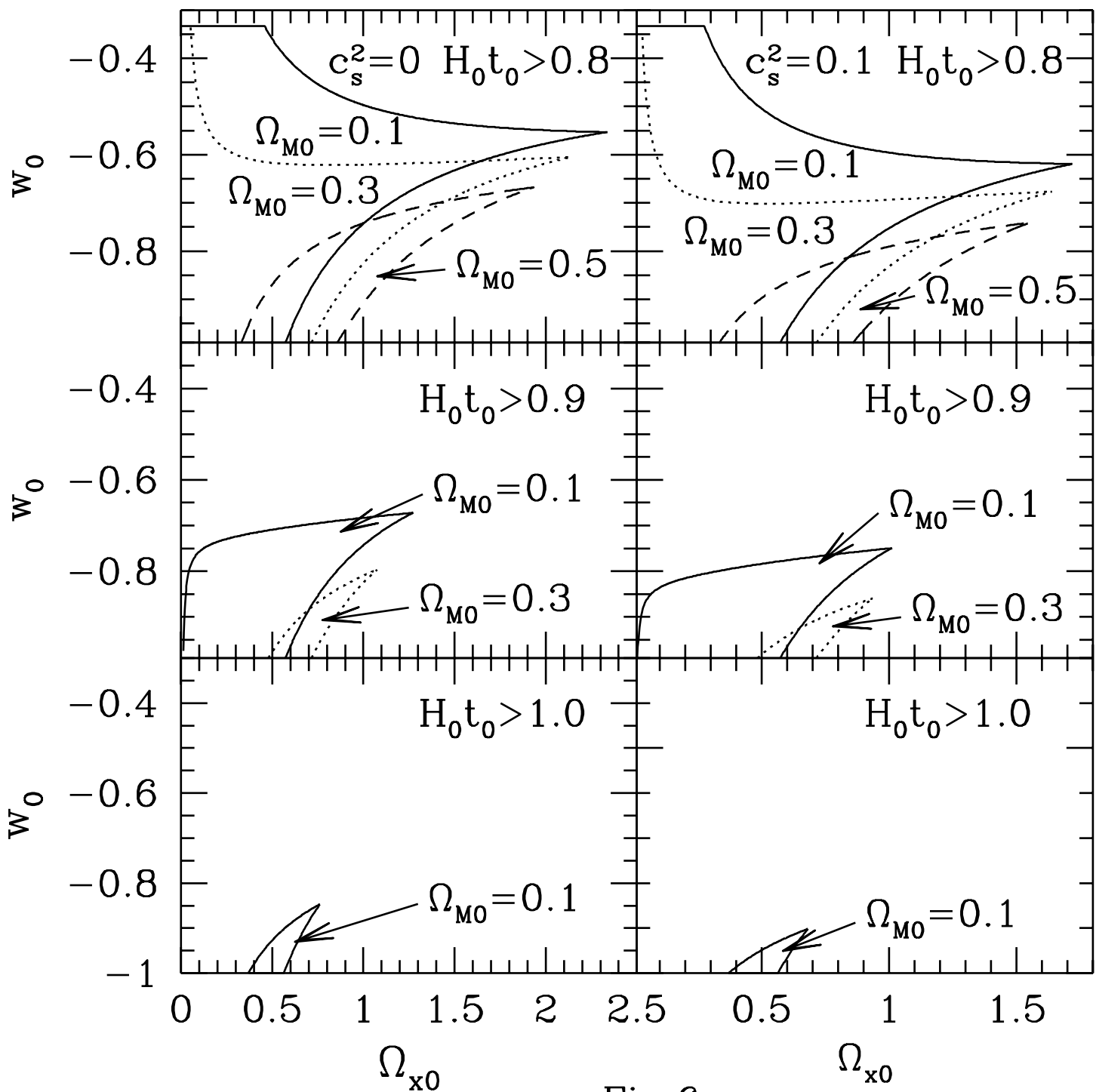
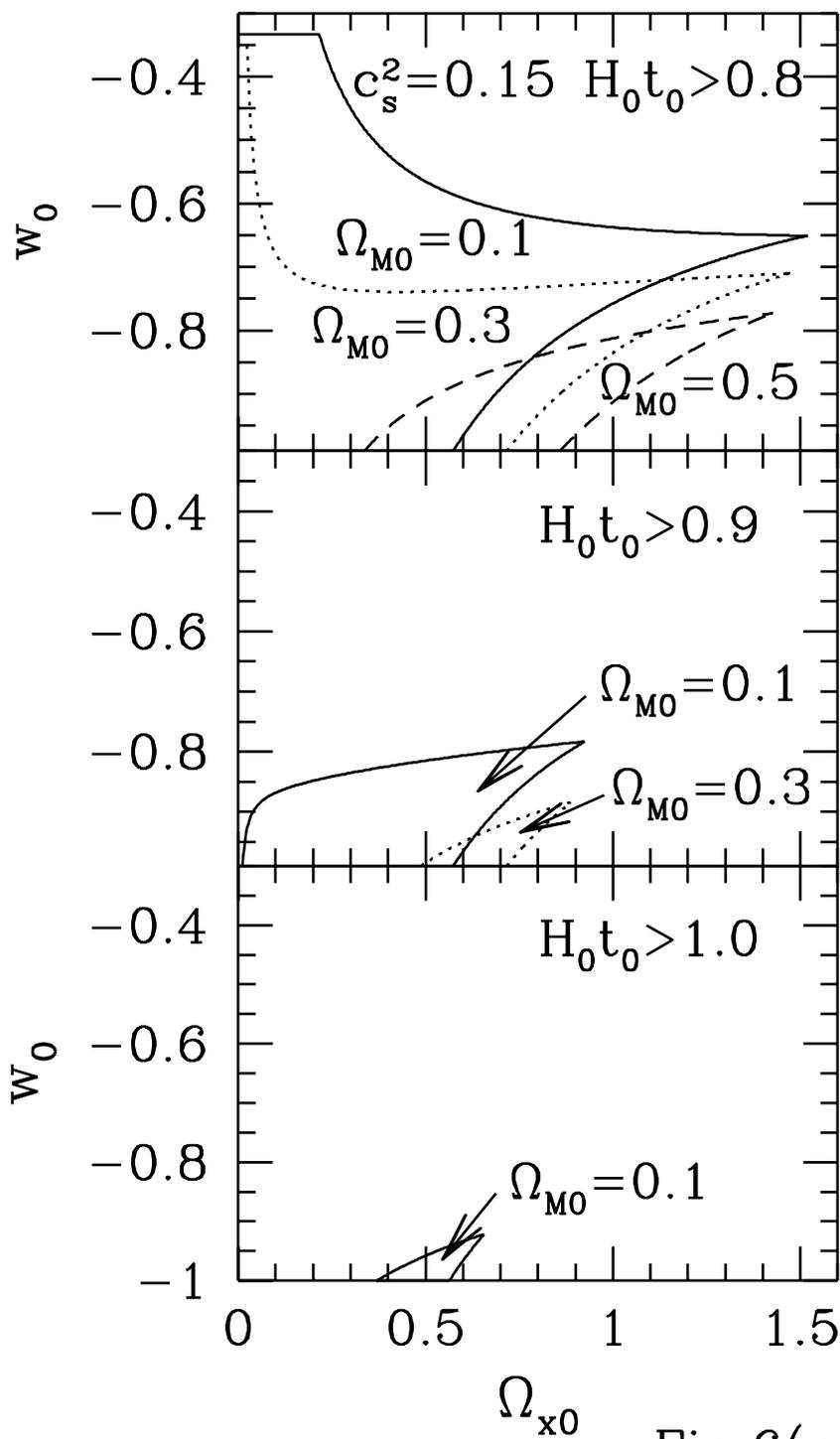
{\bf Figure 6: } Allowed regions of $\Omega_{x0}$ and $w_0$ for 
$\Omega_{M0}=0.1,0.3,0.5$ and $c_s^2=0,0.1,0.15$ 
when  $H_0t_0 > 0.8$, 
$H_0t_0 > 0.9$, or $H_0t_0 > 1.0$. 
A region  bounded by 
a solid line is for  $\Omega_{M0}=0.1$, a dotted line for 
$\Omega_{M0}=0.3$, a dashed line for $\Omega_{M0}=0.5$.

\label{lastpage}

\end{document}